\documentclass[10pt, onecolumn]{article}   	
\usepackage{geometry}                		
\pdfoutput=1                 		
\usepackage{graphicx}
\usepackage{subfigure}				
\usepackage{amssymb}
\usepackage{url}
\usepackage{amsmath}

\newcommand{\argmin}{\operatornamewithlimits{arg\ min}}

\title{\bf Self-Organizing Maps Classification with Application to Laptop's Adapters Magnetic Field}
\author{Darko Brodi\'c$^1$\footnote{(Corresponding author)}, Alessia Amelio$^2$}
\date{\small $^1$ Technical Faculty in Bor, University of Belgrade, Vojske Jugoslavije 12, 19210 Bor, Serbia, \\dbrodic@tfbor.bg.ac.rs\\$^2$Department of Computer Engineering, Modeling, Electronics and Systems, University of Calabria, \\Via P. Bucci Cube 44, 87036 Rende (CS), Italy,\\aamelio@dimes.unical.it}							

\begin{document}
\maketitle

\begin{@twocolumnfalse}

{\bf Abstract}. This paper presents an application of the Self-Organizing-Map classification method, which is used for classification of the extremely low frequency magnetic field emission in the near neighborhood of the laptop adapters. The experiment is performed on different laptop adapters of the same characteristics. After that, the Self-Organizing-Map classification on the obtained emission data is performed. The classification results establish the typical emission levels of the laptop adapters, which are far above the safety standards' limit.  At the end, a discussion is carried out about the importance of using the classification as a possible solution for safely use the laptop adapters in order to reduce the negative effects of the magnetic field emission to the laptop users.
\end{@twocolumnfalse}
\vspace{1cm}

\section{Introduction}
Classification is the process of inferring the category from given input data. There exist two types of classification, according to the objective of the analysis: (i) supervised, and (ii) unsupervised. Supervised classification is based on the process of learning a model. In particular, a training set of data for which the categorization is known is presented to the model. Accordingly, the model parameters are modified to best fit the training data, such that the error between their real categorization and that found by the model is minimized. After training the model, the classifier is ready to categorize new data for which the category is unknown. Unsupervised classification, also known as clustering, is the process of detecting the categories of the input data, which are not prior known. The classification process can be adopted in different context, including data which are generated by some instrumentation measuring specific physical quantities. 

An AC adapter is an electronic device which is essential for each laptop. It is used for supplying the current to the inner electronic components of the laptop as well as to charge the laptop battery, which is used in the laptop to operate flawlessly, when there is no current supply. Still, its operation emits a considerable amount of magnetic field in its neighborhood, especially in the extremely low frequency (ELF) range between 30 and 300 Hz \cite{[1]}. 

Many studies have been made in order to examine the negative consequences of the ELF magnetic field emission to the computer users' health. The IARC monograph has concluded that the ELF magnetic field is possibly carcinogenic to the humans \cite{[2]}. Furthermore, many analyses have shown that the exposure to an ELF magnetic field higher than 0.4 $\mu$T can double the risk of childhood leukemia \cite{[3]}.

The TCO'05 standard dated from 2012 proposed a methodology for magnetic field measurement in the extended extremely low frequency range between 50 Hz and 2 kHz \cite{[4]} and set the safety emission standard reference to 0.2 $\mu$T.

In this paper, we measure the ELF magnetic field emission of 3 different AC adapters made from the same manufacturer of the same characteristics. We measure their ELF emission with the professional measuring device AARONIA NF-5035 \cite{[5]}. Also, we propose a new measurement geometry which takes into account a user-centric measurement approach similar to that given in refs. \cite{[6]}, \cite{[7]}, \cite{[8]}, \cite{[9]}, \cite{[10]}. Then, measured data are classified by the Self-Organizing-Map (SOM) classifier for detecting the dangerous magnetic field emission levels in the neighborhood of a typical laptop's AC adapter.

The paper is organized as follows. Section 2 describes the Self-Organizing Map classification method. Section 3 gives a case study which describes the measurement experiment. After that, it explains the classification experiment. Section 4 presents the measurement results as well as the SOM based classification of them. Finally, Section 5 draws the conclusions.

\section{Self-Organizing Map Classification}

\begin{figure}[!ht]
\begin{center}
\includegraphics[height=10cm, width=10cm, keepaspectratio]{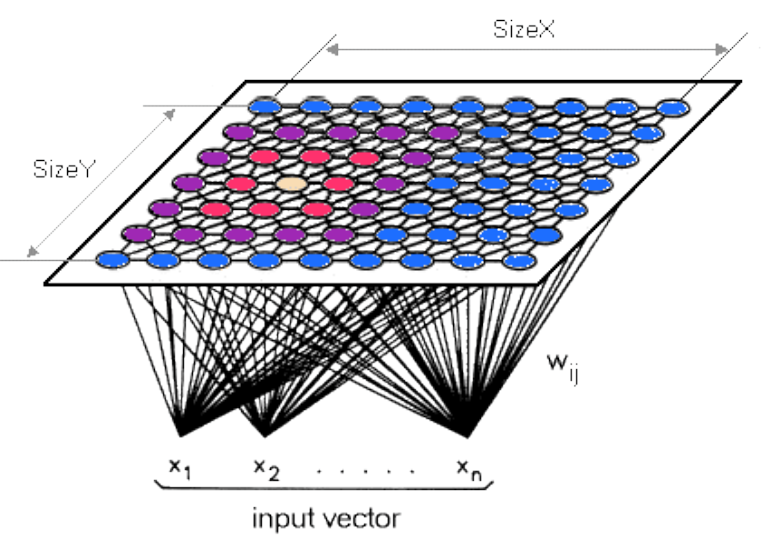}
\caption{An example of SOM network \cite{[12]}}
\label{fig1}
\end{center}
\end{figure}

The Kohonen's Self-Organizing-Map is a classification method which is used for clustering a set of data points according to the creation of a model \cite{[11]}. In particular, the model is an artificial neural network, which is learned to follow the shape of the training data, in order to generate the output. It is characterized by $N$ total neurons, which are positioned on a grid in one or two dimensions. The $K$ output neurons are connected by following a neighborhood typology. In order to realize the clustering, the number of the output neurons can be set as the number of output clusters. The $n$ input neurons correspond to the $n$ dimensions of an input data point $x = \{x_1 x_2 ... x_n\}$. An output neuron $i$ is associated with a prototype vector, $w_i = \{w_{i1} w_{i2} ... w_{in}\}$, which is a $n$-dimensional vector of weights between $i$ and the $n$ input neurons. Each input neuron is connected to all the output neurons. Figure \ref{fig1} shows the network with $n$ input neurons and $K$ output neurons.

The algorithm for training the network iterates over different epoches ($t = 1,...,T$) on the training data until computational bounds do not exceed. The first step consists of initializing all the weights to small random values. In the current iteration, the algorithm randomly chooses an input data point $x$ from the training set to be processed by the network. The output neuron $i^*$ with the closest prototype vector in terms of Euclidean distance to the input data point $x$ is selected as the Best-Matching Unit (BMU):
	 	
\begin{equation}
i^*= \argmin_i ||x - w_i||^2.
\end{equation}

All the prototype vectors in the neighborhood of the BMU are updated by using the following update formula:

\begin{equation}
w_i(t+1)=w_i(t) + \eta(t) (x-w_i(t)),
\end{equation}	 	
where $\eta$ is the learning rate, which monotonically decreases over time. At the end of the learning process, the network is ready to classify the test input data points. In particular, the data point is presented as the input to the network, and its BMU will correspond to its output class (or cluster). The algorithm takes $O(N^2)$ for its execution, where $N$ is the total number of neurons in the map. 

The main advantage of the SOM is the dimensionality reduction of the input data into a one or two dimensional grid. It makes easier to understand and interpret the data. Another important advantage is that SOM is able to manage a large variety of input data because it adapts to their shape. Also, the SOM is able to cluster complex and large data sets in a reduced time. Finally, the SOM usually works fine, being able to correctly find the classification of the data points. This aspect makes the SOM an invaluable instrument to be used in multiple contexts and domains where the traditional classification methods fail to correctly classify the input data. 

\section{A Case Study: Self-Organizing-Map for Classification of Laptop's AC Adapters Magnetic Field}

\subsection{Measurement Experiment}

\begin{figure}[!ht]
\begin{center}
\subfigure[]{
\includegraphics[height=7.5cm, width=7.5cm, keepaspectratio]{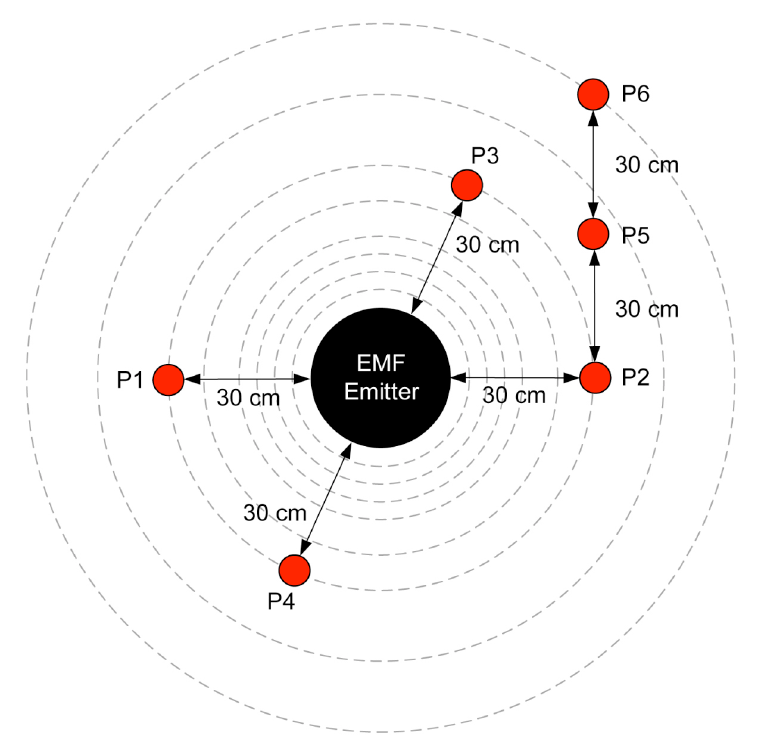}}
\subfigure[]{
\includegraphics[height=8.5cm, width=8.5cm, keepaspectratio]{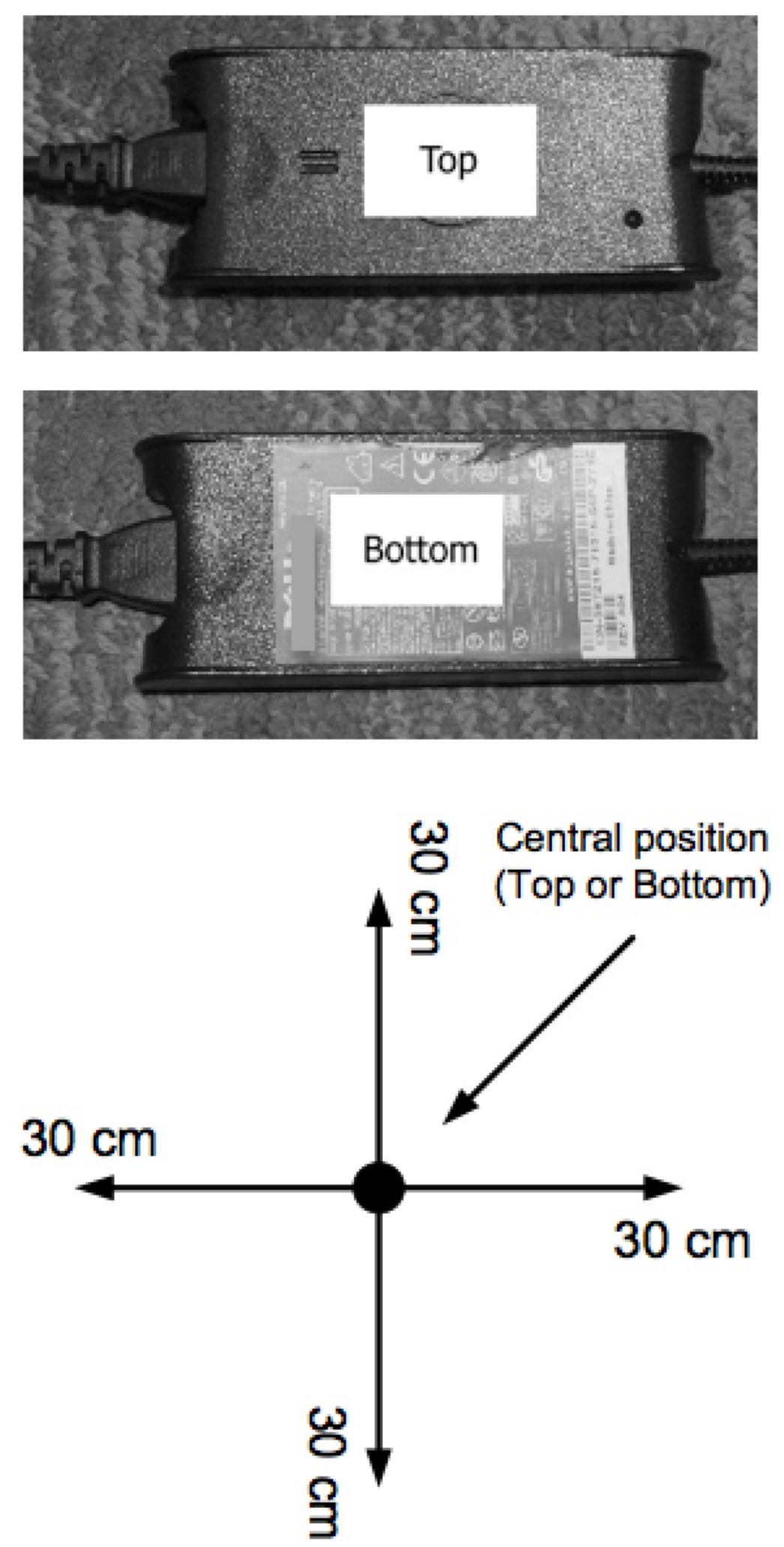}}
\caption{The measurement geometry: (a) the proposed TCO, (b) our measurement geometry}
\label{fig2}
\end{center}
\end{figure}

The measurement experiment includes the ELF magnetic field measurement of 3 different AC adapters of the same power equal to 90 W from the same manufacturer, designated as AC1, AC2 and AC3. These AC adapters are used with a laptop of the same manufacturer, which is operated in the so-called office mode using the operating system Microsoft Windows 7. It means that the laptop is used for Internet browsing, YouTube, Office calculation, typing documents, etc. Hence, the measurement experiment consists of measuring the ELF magnetic field that the AC adapters emit in their closest neighborhood. The measurement geometry is an extension of the TCO proposed one. The TCO measurement geometry proposes six measurement positions which are 30 cm in each of four directions (top, down, left and right) away from the emitter as well as two additional measurement positions 30 cm away in the direction down, but 30 and 60 cm up \cite{[4]}. We slightly change this proposed geometry. Basically, we measure the ELF magnetic field at the top and bottom of the AC adapter as well as at the 4 positions proposed by TCO, i.e. 30 cm in each of four directions (top, down, left and right) away from the emitter. Differently from the TCO standard, our measurement geometry takes into account the real proximity of the user to the laptop's AC adapter which occurs when he/she touches it on its top or bottom side by using the hands or the fingers. Figure \ref{fig2} illustrates the TCO measurement geometry, as well as our new proposed measurement geometry.

The ELF magnetic field is measured with the measuring device Aaronia Spectran NF-5035 \cite{[5]}. It has a frequency range from 1 Hz to 1 MHz. In our case, we only use the mentioned ELF range from 30 to 300 Hz. This device has the capability of measuring the minimum, maximum, average and true RMS (Root Mean Square) magnetic induction $B$ in the range from 1 pT to 2 mT, with a sampling time of at least 10 ms. We use it to measure each scalar component of the magnetic induction $B$, i.e. $B_x$, $B_y$ and $B_z$ as well as the magnitude of the magnetic induction $B$ in the ELF range.

\subsection{Classification Experiment}
The SOM method is applied on the ELF magnetic field data of the 3 tested laptop's AC adapters for detecting the magnetic field levels emitted in the neighborhood of a typical laptop's AC adapter. In particular, we investigate the ability of the classifier in finding the emission levels at different frequency bands to which the users are subjected during their contact with a laptop's AC adapter. This analysis is of particular interest, because multiple tests with traditional methods for data discretization (e.g. Equal-Width or Frequency binning) failed in correctly characterizing the magnetic field levels. 

\begin{figure}[!ht]
\begin{center}
\includegraphics[height=10cm, width=10cm, keepaspectratio]{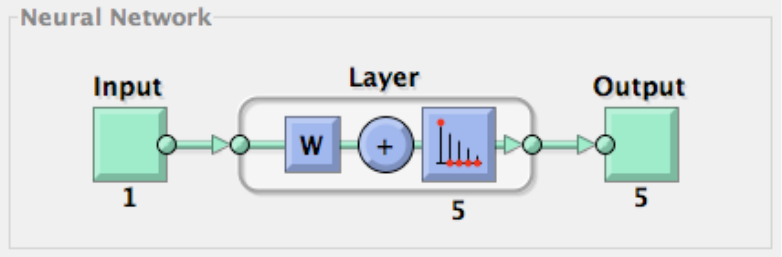}
\caption{The Kohonen's network adopted in the experiment}
\label{fig3}
\end{center}
\end{figure}

In particular, we create 6 distinct datasets, each containing the measured data of the 3 AC adapters in the overall frequency range between 30 and 300 Hz at intervals of 5-6 Hz in one side of the AC adapters: (i) top, (ii) bottom, (iii) 30 cm down from the adapter, (iv) 30 cm up to the adapter, (v) 30 cm from the left of the adapter, and (vi) 30 cm from the right of the adapter. Because the magnetic field data are measured at different time points for each frequency, we compute the average over the time interval, in order to obtain one value for each frequency. Accordingly, each dataset is characterized by 153 instances (51 instances $\times$ 3 adapters), each represented by one feature, which is the measured magnetic field value in one side of an AC adapter at a given frequency. Each dataset is processed by the SOM method which partitions the instances in different clusters, corresponding to the emission levels in the different frequency bands. Because of the one-dimensional instances, the Kohonen's network is composed of one input neuron. Also, the number of output neurons is set to 5, in order to obtain 5 emission levels of the magnetic field: (i) very low, (ii) low, (iii) middle, (iv) high, and (v) very high. Figure \ref{fig3} shows the Kohonen's network adopted for the experiment.

In the network, the number of training steps for initial covering of the input space is 100 and the initial size of the neighborhood is 3. The distance between two neurons is calculated as the number of steps dividing one from the other.

\section{Results and Discussion}
\subsection{Measurement Results}

The measurement results of the ELF magnetic field emitted by the tested AC adapters are given in Table \ref{tab1}.

\begin{table}[!ht]
\caption{The ELF magnetic field emission of the AC adapters}
\begin{center}
\scriptsize
\begin{tabular}{llllllll}
No.&	AC Adapter&	$B_{Top}$ [$\mu$T]&	$B_{Bottom}$ [$\mu$T]&	$B_{Left}$ [$\mu$T]&	$B_{Right}$ [$\mu$T]&	$B_{Up}$ [$\mu$T]&	$B_{Down}$ [$\mu$T]\\\\\hline\\
AC1	&Adapter No. 1&	4.910&	4.220&	0.350&	0.348&	0.348&	0.348\\\\
AC2	&Adapter No. 2&	4.360&	4.210&	0.350&	0.348&	0.364&	0.345\\\\
AC3&	Adapter No. 3&	15.400&	4.260&	0.460&	0.462&	0.452&	0.451\\\\
\hline
\end{tabular}
\end{center}
\label{tab1}
\end{table}%

\begin{figure}[!ht]
\begin{center}
\subfigure{
\includegraphics[height=6.5cm, width=6.5cm, keepaspectratio]{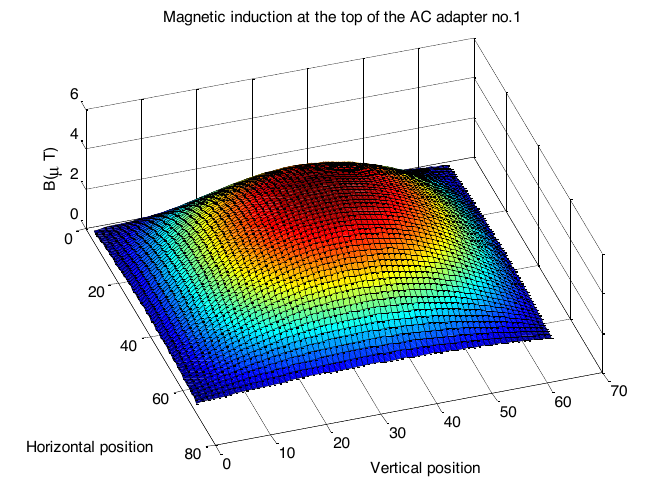}}
\subfigure{
\includegraphics[height=6.5cm, width=6.5cm, keepaspectratio]{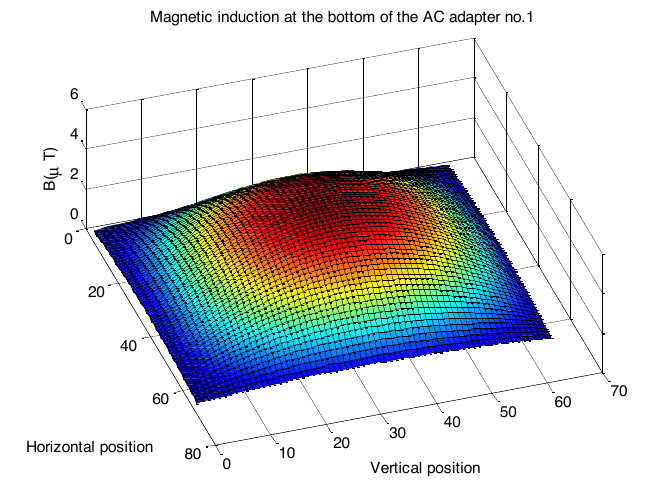}}
\subfigure{
\includegraphics[height=6.5cm, width=6.5cm, keepaspectratio]{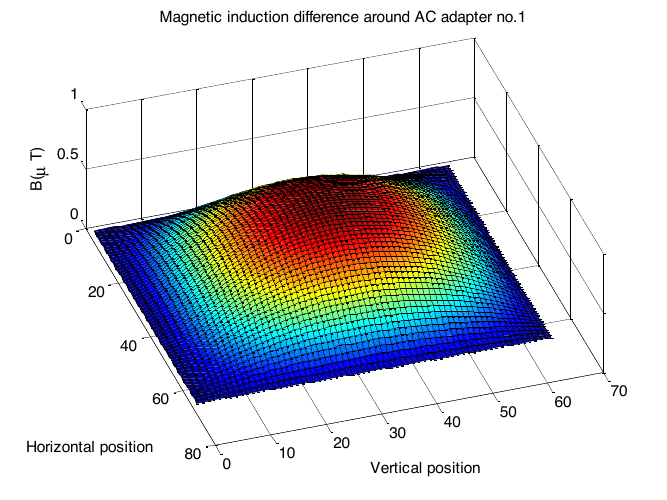}}
\caption{AC1 magnetic field: at the top (left), at the bottom (right), the difference (bottom)}
\label{fig4}
\end{center}
\end{figure}

\begin{figure}[!ht]
\begin{center}
\subfigure{
\includegraphics[height=6.5cm, width=6.5cm, keepaspectratio]{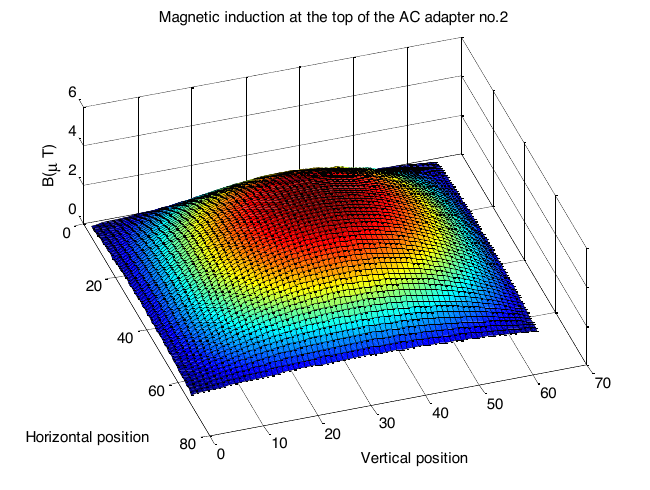}}
\subfigure{
\includegraphics[height=6.5cm, width=6.5cm, keepaspectratio]{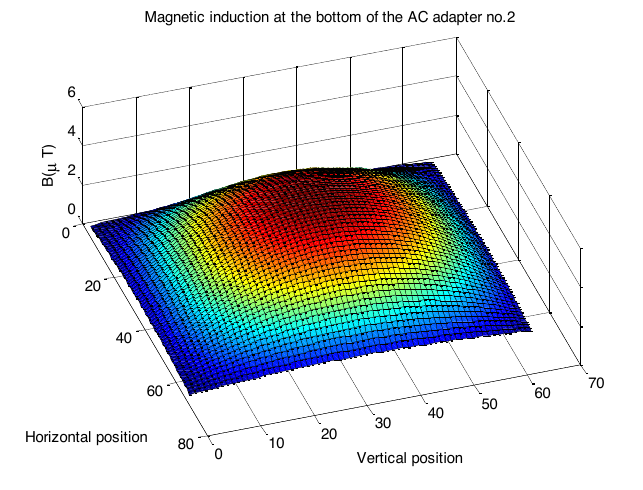}}
\subfigure{
\includegraphics[height=6.5cm, width=6.5cm, keepaspectratio]{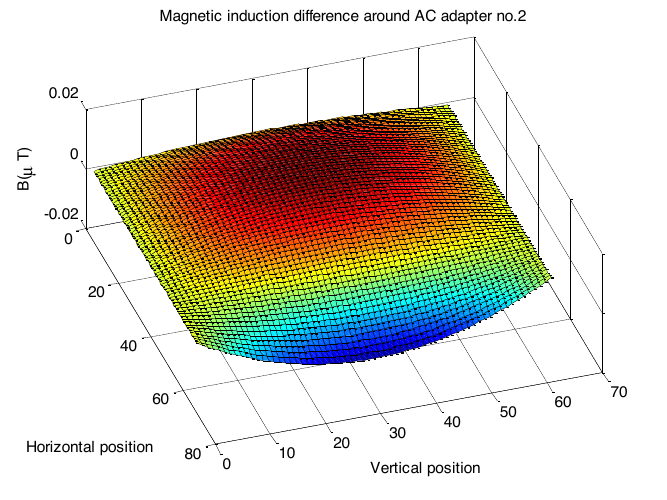}}
\caption{AC2 magnetic field: at the top (left), at the bottom (right), the difference (bottom)}
\label{fig5}
\end{center}
\end{figure}

\begin{figure}[!ht]
\begin{center}
\subfigure{
\includegraphics[height=6.5cm, width=6.5cm, keepaspectratio]{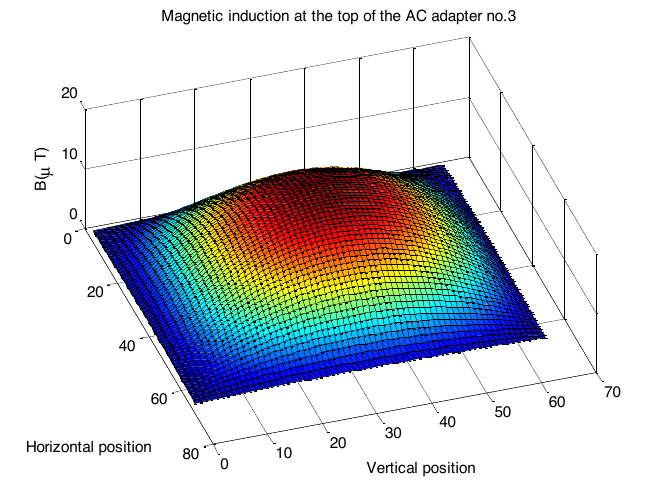}}
\subfigure{
\includegraphics[height=6.5cm, width=6.5cm, keepaspectratio]{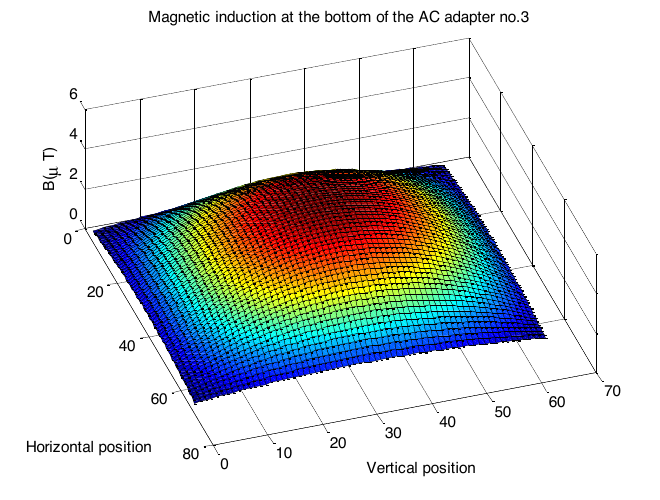}}
\subfigure{
\includegraphics[height=6.5cm, width=6.5cm, keepaspectratio]{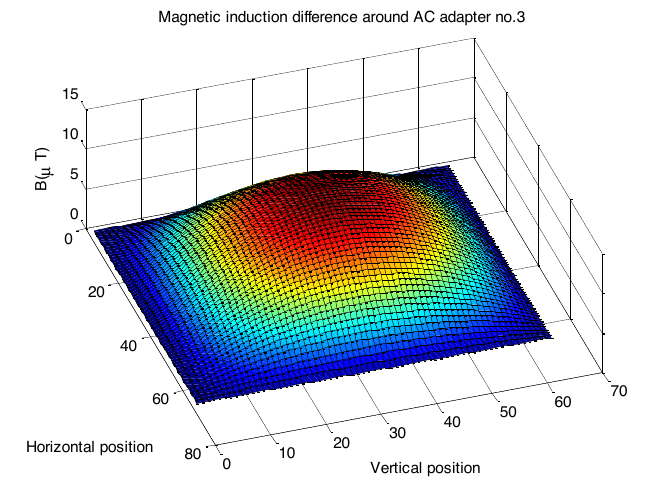}}
\caption{AC3 magnetic field: at the top (left), at the bottom (right), the difference (bottom)}
\label{fig6}
\end{center}
\end{figure}

From Table \ref{tab1}, it is visible that all tested adapters emit an amount of ELF magnetic field in their normal office working condition which is much higher than the safety reference limit given by the TCO standard. It is worth noting that their emission level is also above the safety limits at the distance of 30 cm away from the emitter, i.e. AC adapter \cite{[4]}. Figures \ref{fig4}-\ref{fig6} graphically illustrate their emission at: (i) the top part, (ii) the bottom part, and (iii) the difference between top and bottom parts. 

Also, it is interesting to observe that AC3 emits a much higher level of ELF magnetic field in a similar working condition (used with the same laptop in the same working conditions). Hence, it can be said that AC1 and AC2 are better engineered than AC3. Furthermore, the top part of the AC adapter should emit a lower level of ELF magnetic field than its bottom part. If we recall the results given in Table \ref{tab1}, it is quite clear that all three AC adapters do not satisfy such criteria. However, the AC3 adapter is the worst among them emitting an ELF magnetic field which is approximately 3.5$\times$ higher at its top part. From all the aforementioned, it is quite clear that AC adapters emit a high level of ELF magnetic field and that laptop users should be staying away from them as much as possible. 

\subsection{Classification Results}
\begin{figure}[!ht]
\begin{center}
\includegraphics[height=15.5cm, width=15.5cm, keepaspectratio]{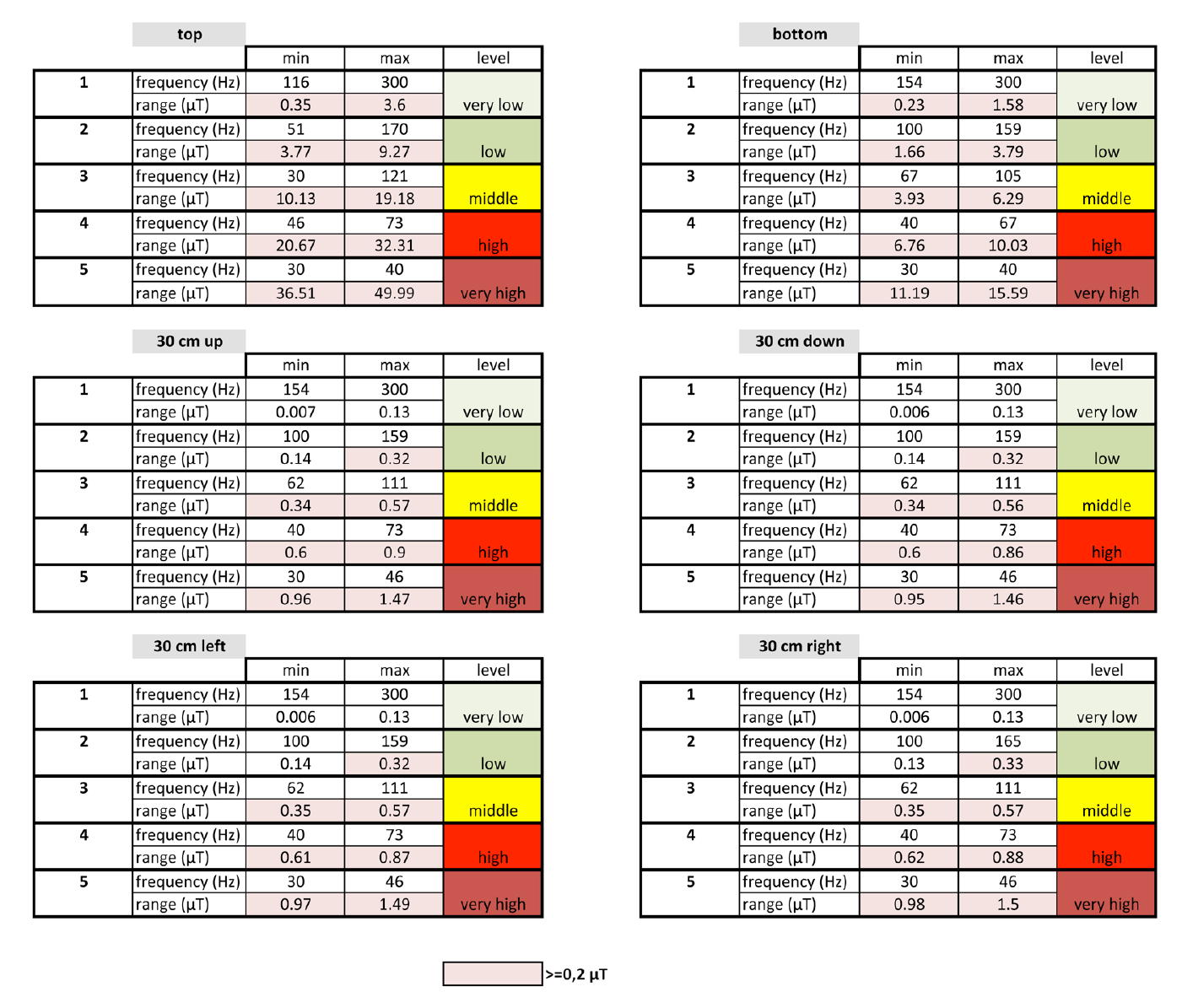}
\caption{ELF magnetic field levels obtained by the SOM method on the 6 datasets}
\label{fig7}
\end{center}
\end{figure}

The results of the experiment using the SOM method for clustering the 6 datasets are shown in Figure \ref{fig7}. Each table illustrates the 5 obtained ELF magnetic field levels emitted at the different sides of a typical laptop's AC adapter. Each level is differently colored and represented by its minimum and maximum values of magnetic field and frequency. Values overcoming the safety reference limit of 0.2 $\mu$T are marked in pink. A first important observation concerns the emission level, which is the highest at the top side of the adapter, reaching a peak between 36.51 and 49.99 $\mu$T in the very high level. It is followed by the bottom side, ranging between 11.19 and 15.59 $\mu$T in the very high range. The other sides never overcome 1.5 $\mu$T, which is much lower than the top and the bottom. In any case, it is observed that most of the emission levels are far above the reference limit of 0.2 $\mu$T in all sides of the adapter. Only a few exceptions are visible at 30 cm up, down, on the left and on the right of the adapter, with very low and low ranges below the reference limit of 0.2 $\mu$T or borderline (minimum value below and maximum value above). It does not represent an encouraging news, considering that users are often in close contact with the laptop's AC adapter with their hands and fingers when they touch it. Hence, the distance of 30 cm away from the AC adapter cannot be always guaranteed. This represents a clear limitation of the TCO standard measurement geometry. Another interesting observation is that the highest emission levels are always associated with the lowest frequency bands. Specifically, the most dangerous magnetic field levels (high and very high) at the top and bottom side of the AC adapter are emitted in a frequency band between 30 and 70 Hz. A similar condition can be observed at 30 cm up, down, on the left and on the right of the AC adapter, where the high and very high levels correspond to a frequency band of 30-73 Hz. On the contrary, the lowest emission levels are associated with the highest frequency ranges between 150 and 300 Hz (see very low level). 

Hence, we can observe that the laptop's AC adapter emits at ELF magnetic field levels which are prohibitive for the human health, according to the safety reference limit. Also, it emits the strongest magnetic field in correspondence of the lowest frequencies. 

According to the SOM classification results, in order to reduce the potential risk connected with the use of the laptop's AC adapter, some easy rules should be followed by the user: (i) to avoid the direct contact of the user's hands and fingers with the adapter, (ii) because the magnetic field emission is reduced by the distance, to keep the adapter far from the body (more than 30 cm away), (iii) because the strongest emission is registered in correspondence of one side of the adapter, it is recommended to keep this side on the table or the desk. Consequently, the SOM method revealed to be a very important and useful approach for safely using the laptop's AC adapters. In fact, it enriched the traditional measurement process by introducing a range division of the magnetic field emission into typical dangerousness levels, which more easily show the potential risk connected with the use of the AC adapter.

\section{Conclusion}
The presented analysis introduced a case study of classification of the measured ELF magnetic field data around a laptop's AC adapter by using the Self-Organizing Map. The classification proved the potential danger risk of using the AC adapter, when the laptop users are in its close neighborhood, which is many times higher than the safety reference limit. At the end, some practical proposition is given in order to reduce the users' exposure to the high level of ELF magnetic field emitted by the AC adapter. It revealed to be a very useful approach in the state-of-the-art to evaluate and prevent the potential hazardous effects of the ELF magnetic field exposure to the human's health. 
\\\\\\
{\bf Acknowledgement.}
This study was partially funded by the Grant of the Ministry of Education, Science and Technological Development of the Republic of Serbia, as a part of the project TR33037.

\end{document}